\documentclass[sigconf]{acmart} 
\usepackage{booktabs} 
\usepackage{amsmath}
\usepackage{amsfonts}
\usepackage{comment}
\usepackage{footmisc}
\usepackage{mathrsfs}
\usepackage{amssymb}
\usepackage{graphicx}
\usepackage{subfigure}
\usepackage{multirow}
\usepackage{algorithm}
\usepackage{algorithmic}
\usepackage{multicol}  
\usepackage{enumitem}
\usepackage{array}
\usepackage{float}
\usepackage{hyperref}
\AtBeginDocument{%
	\providecommand\BibTeX{{%
			\normalfont B\kern-0.5em{\scshape i\kern-0.25em b}\kern-0.8em\TeX}}}

\copyrightyear{2020}
\acmYear{2020}
\setcopyright{acmcopyright}
\acmConference[CIKM '20]{Proceedings of the 29th ACM International Conference on Information and Knowledge Management}{October 19--23, 2020}{Virtual Event, Ireland}
\acmBooktitle{Proceedings of the 29th ACM International Conference on Information and Knowledge Management (CIKM '20), October 19--23, 2020, Virtual Event, Ireland} \acmPrice{15.00}
\acmDOI{10.1145/3340531.3412044}
\acmISBN{978-1-4503-6859-9/20/10}

\begin{document}
\fancyhead{}

\title{Whole-Chain Recommendations}
\author{Xiangyu Zhao}
\affiliation{
	\institution{Michigan State University}
}
\email{zhaoxi35@msu.edu}

\author{Long Xia}
\affiliation{
	\institution{York University}
}
\email{longxia@yorku.ca}

\author{Lixin Zou}
\affiliation{
	\institution{Baidu Inc.}
}
\email{zoulixin15@gmail.com}

\author{Hui Liu}
\affiliation{
	\institution{Michigan State University}
}
\email{liuhui7@msu.edu}

\author{Dawei Yin}
\affiliation{
	\institution{Baidu Inc.}
}
\email{yindawei@acm.org}

\author{Jiliang Tang}
\affiliation{
	\institution{Michigan State University}
}
\email{tangjili@msu.edu}
\renewcommand{\shortauthors}{Xiangyu Zhao et al.}

\begin{abstract}
	With the recent prevalence of Reinforcement Learning (RL), there have been tremendous interests in developing RL-based recommender systems. In practical recommendation sessions, users will sequentially access multiple scenarios, such as the entrance pages and the item detail pages, and each scenario has its specific characteristics. However, the majority of existing RL-based recommender systems focus on optimizing one strategy for all scenarios or separately optimizing each strategy, which could lead to sub-optimal overall performance. In this paper, we study the recommendation problem with multiple (consecutive) scenarios, i.e., whole-chain recommendations. We propose a multi-agent RL-based approach (DeepChain), which can capture the sequential correlation among different scenarios and jointly optimize multiple recommendation strategies. To be specific, all recommender agents (RAs) share the same memory of users' historical behaviors, and they work collaboratively to maximize the overall reward of a session. Note that optimizing multiple recommendation strategies jointly faces two challenges in the existing model-free RL model~\cite{lowe2017multi} - (i) it requires huge amounts of user behavior data, and (ii) the distribution of reward (users' feedback) are extremely unbalanced. In this paper, we introduce model-based RL techniques to reduce the training data requirement and execute more accurate strategy updates. The experimental results based on a real e-commerce platform demonstrate the effectiveness of the proposed framework. 
\end{abstract}
\maketitle
\section{Introduction}
\label{sec:introduction}
With the recent tremendous development in Reinforcement Learning (RL), there have been increasing interests in adapting RL for recommendations. RL-based recommender systems treat the recommendation procedures as sequential interactions between users and a recommender agent (RA). They aim to automatically learn an optimal recommendation strategy (policy) that maximizes cumulative rewards from users without any specific instructions. RL-based recommender systems can achieve two key advantages: (i) the RA can learn their recommendation strategies based on users' real-time feedback during the user-agent interactions continuously; and (ii) the optimal strategies target at maximizing the long-term reward from users (e.g. the overall revenue of a recommendation session). Therefore, numerous efforts have been made on developing RL-based recommender systems~\cite{zhao2018recommendations,zhao2017deep,zhao2018deep,dulac2015deep,fan2020attacking,feng2018learning}. 

\begin{figure*}[t]
	\centering
	\includegraphics[width=\linewidth]{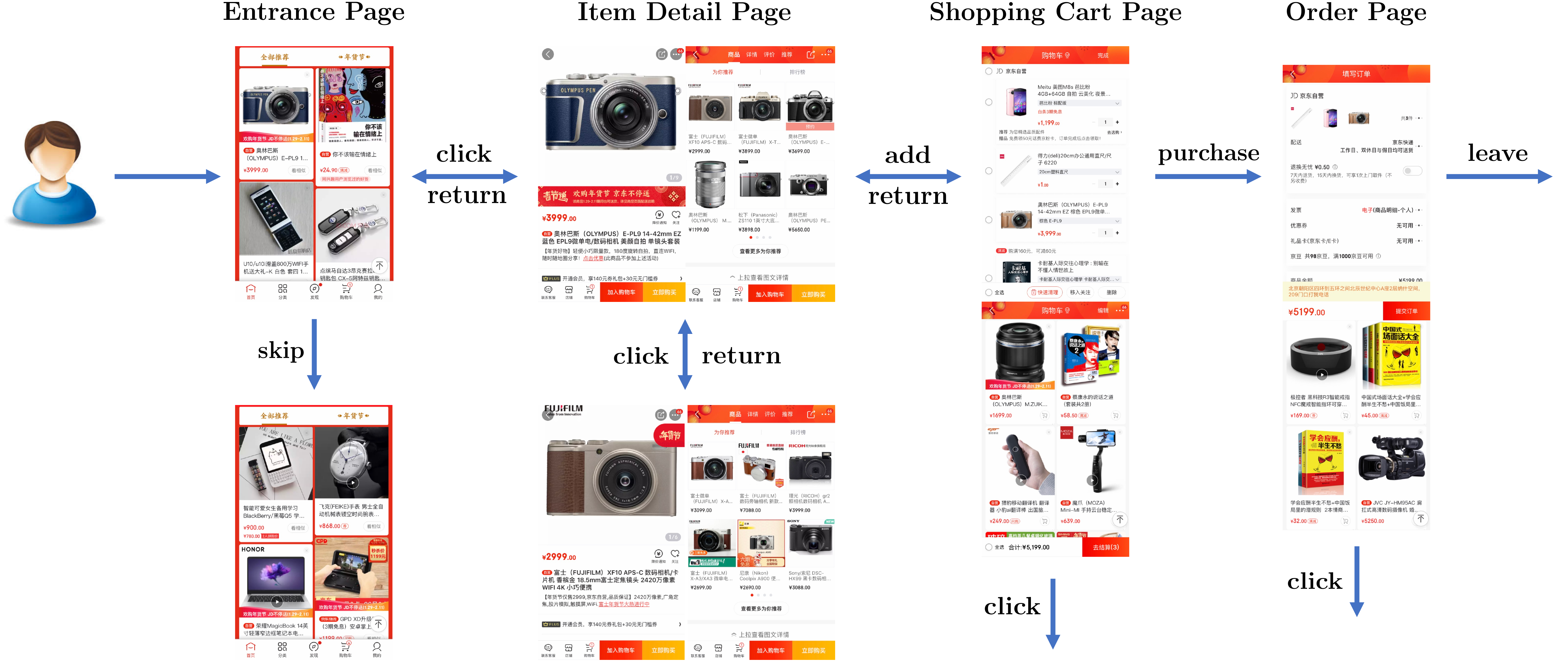}
	\caption{An example of whole-chain recommendations.}
	\label{fig:mdp_example}
\end{figure*}

In reality, as shown in Figure~\ref{fig:mdp_example}, users often sequentially interact with multiple RAs from different scenarios in one recommendation session. First, a user usually starts a recommendation session by browsing the recommended items in the \textit{entrance page} of the E-commerce platform, which suggests diverse and complementary items according to the user's browsing history where the user can: (i) skip the recommended items and continue browsing the new recommendations, or (ii) go to the \textit{item detail page} if she clicks one preferred item. Second, the \textit{item detail page} shows the details of the clicked item, and the RA of this page recommends a set of items related to the clicked item where the user can (i) go back to the entrance page, (ii) go to another item detail page if she clicks one recommended item, or (iii) add the item into the shopping cart and go to the \textit{shopping cart page}. Third, the \textit{shopping cart page} lists all items that the user have added, and an RA generates recommendations associated with these items where the user can (i) return to the last item detail page, (ii) click one recommended item and go the item detail page, or (iii) go to the \textit{order page} if she decides to purchase some items. Finally, after purchasing items in the \textit{order page}, an RA will recommend a set of items related to the purchased items. Note that (i) the user will be navigated to an item detail page wherever she clicks a recommended item, and (ii) the user can leave the platform at any scenarios (we only show one ``leave'' behavior in Figure~\ref{fig:mdp_example}). 

The real example suggests that there is a chain of recommendation scenarios and these scenarios are sequentially related. However, the majority of traditional methods usually only optimize one recommendation strategy for all scenarios or independently optimize each recommendation strategy based on the data from each scenario~\cite{zhao2020memory,zhao2019ads,zou2020neural,zhao2019toward,zhao2020jointly,zhao2020autoemb,zhao2018recommendations,zhao2017deep,zhao2018deep,zhao2019deep,zheng2018drn,zou2020pseudo,zou2019reinforcement}, which could result in sub-optimal overall performance. First, from the above example, different scenarios have independent objectives, e.g., the \textit{entrance page} focuses on the trade-off between correlation and diversity while the \textit{item detail page} focuses more on correlation, thus optimizing only one strategy for all scenarios is sub-optimal. Second, separate optimization ignores the sequential correlation and dependency of users' behaviors among different scenarios. Third, optimizing one strategy within a specific scenario only leverages the user-agent interaction data within this scenario, while completely ignoring the information (users' behaviors) from other scenarios. Finally, independent optimization of one scenario only maximizes its objective, which may negatively affect the overall objective of the whole recommendation session. In other words, recommending an item in one specific scenario may negatively influence user's click/purchase behaviors in other scenarios. Thus, in this paper, we formulate the recommendation tasks within multiple consecutive scenarios as a whole-chain recommendation problem, and leverage multi-agent reinforcement learning (MARL) to jointly optimize multiple recommendation strategies, which is capable of maximizing the overall performance of the whole recommendation session. The designed whole-chain recommendation framework (DeepChain) has three advantages. First, RAs are sequentially activated to capture the sequential dependency of users' behaviors among different scenarios. Second, all RAs in different scenarios share the same memory of historical user behavior data, in other words, an RA in one scenario can leverage user behavior data from other scenarios to make more accurate decisions. Third, all RAs can jointly work to optimize the overall performance of the whole recommendation session.

In order to optimize recommendation strategies, existing model-free RL-based recommender systems typically require a larger amount of user-agent interaction data~\cite{dulac2015deep,zhao2018recommendations,zheng2018drn,feng2018learning}. The whole-chain setting with multiple scenarios demands even more data. However, this requirement is challenging in practical recommendation systems, since real-world users will leave the platforms quickly if the systems randomly recommend items that cannot fit users' preferences~\cite{chen2018neural}. Furthermore, the distributions of users' immediate feedback (reward) on the recommended items are extremely unbalanced in users' historical logs, since users' click/purchase behaviors (with positive reward) occur much infrequently than users' skip behaviors (with zero reward). This will lead to an inaccurate update of the action-value function of RL~\cite{hu2018reinforcement}. Therefore, to tackle these challenges, in this paper, we propose a model-based RL framework for the MARL-based recommender systems (DeepChain). Compared with model-free models (e.g. \cite{feng2018learning}), the model-based framework approximates the user behaviors (environment) to reduce training data amount requirement and performs accurate optimization of the action-value function. We demonstrate the effectiveness of the proposed framework on a real-world dataset from an e-commerce platform JD.com, and validate the importance of the components in DeepChain for accurate recommendations.
\section{Problem Statement}
\label{sec:problem} 
We formulate the whole-chain recommendation task as a multi-agent model-based RL problem. To be specific, there exist several RAs corresponding to different recommendation scenarios. Each RA serves as a recommendation strategy that recommends items to a user (the environment $\mathcal{E}$) in a specific scenario. Furthermore, the RAs sequentially interact with the user by recommending items over a sequence of time steps, thus the RAs are sequentially activated according to the user's behaviors, and only one RA is activated at each time step. All RAs work cooperatively to maximize the cumulative reward of a recommendation session. In this paper, we model the above multi-agent model-based RL problem as a Markov Decision Process (MDP), which contains a sequence of states, actions and rewards. Formally, the MDP is a tuple with five elements $(\mathcal{S}, \mathcal{A}, \mathcal{P}, \mathcal{R}, \gamma)$ as:

\begin{itemize}[leftmargin=*]
	\item {\bf State space $\mathcal{S}$}: The state $s_t \in \mathcal{S}$ is defined as a chronologically sorted sequence of a user's historical clicked or purchased items before time $t$, which represents the user's preference at time $t$.
	\item {\bf Action space $\mathcal{A}$}:  An action $a_t \in \mathcal{A}$ of an RA is recommending a list of relevant items corresponding to state $s_t$. Without the loss of generality, in this paper, each time an RA recommends only one item to the user, while it is straightforward to extend the setting to recommend multiple items.  
	\item {\bf Reward $\mathcal{R} (\mathcal{S} \times \mathcal{A} \rightarrow \mathbb{R})$}: When an RA recommends an item to a user at time $t$ (i.e. taking action $a_t$), the user will browse the recommended item and provide corresponding feedback (such as skip, click, purchase or leave), and then the RA will receive an immediate reward $r(s_t,a_t)$ based on the user's feedback.
	\item {\bf Transition probability $\mathcal{P} (\mathcal{S} \times \mathcal{A} \times \mathcal{S} \rightarrow [0,1])$}: Transition probability $p(s_{t+1}|s_t,a_t)$ is defined as the probability of state transiting from $s_t$ to $s_{t+1}$ when action $a_t$ is executed by an RA. The MDP is assumed to satisfy the Markov property $p(s_{t+1}|s_t,a_t,...,s_1,a_1) = p(s_{t+1}|s_t,a_t)$. In our setting, the transition probability is equivalent to user behavior probability, which is also associated with the activation of RAs.
	\item {\bf Discount factor $\gamma$}: the reward discount factor $\gamma \in [0,1]$ is leveraged to calculate the present value of future reward. When $\gamma = 1$, all future rewards can be fully counted into the current action; when $\gamma = 0$, only the immediate reward is considered.
\end{itemize}

With the aforementioned definitions and descriptions, we formally define the whole-chain recommendation problem as follows: \textit{Given the historical MDP, i.e., $(\mathcal{S}, \mathcal{A}, \mathcal{P}, \mathcal{R}, \gamma)$, the goal is to find a set recommendation policies $\{\pi\}:\mathcal{S} \rightarrow \mathcal{A}$ for multiple recommender agents of different recommendation scenarios, which can maximize the cumulative reward of the whole recommendation session.}

\section{The Proposed Framework}
\label{sec:framework}
In this section, we will propose a deep RL approach for the whole-chain recommendation problem, which can simultaneously learn multiple recommendation strategies for different scenarios by a model-based RL algorithm. As discussed in Section~\ref{sec:introduction}, developing a whole-chain recommendation framework is challenging, because (i) optimizing only one strategy for all scenarios overlooks the different objectives of different scenarios, (ii) optimizing each strategy for each scenario separately neglects the sequential correlation among scenarios and the information from other scenarios, and solely optimizes its objective may lead to the sub-optimal overall performance of the whole session, and (iii) jointly optimizing multiple recommendation strategies requires substantial user behavior data, and the users' feedback (reward) distributions are extremely unbalanced. To address these challenges, we propose a multi-agent model-based RL framework. Note that for the sake of simplicity, we will only discuss the recommendations within two scenarios, i.e., entrance page and item detail page, however, it is straightforward to extend the setting with more scenarios. In the following, we will first illustrate the overview of the proposed framework, then introduce the architectures of RAs (actors) and critic separately, and finally we will discuss the objective function with its optimization.

\begin{figure}[t]
	\centering
	\includegraphics[width=81mm]{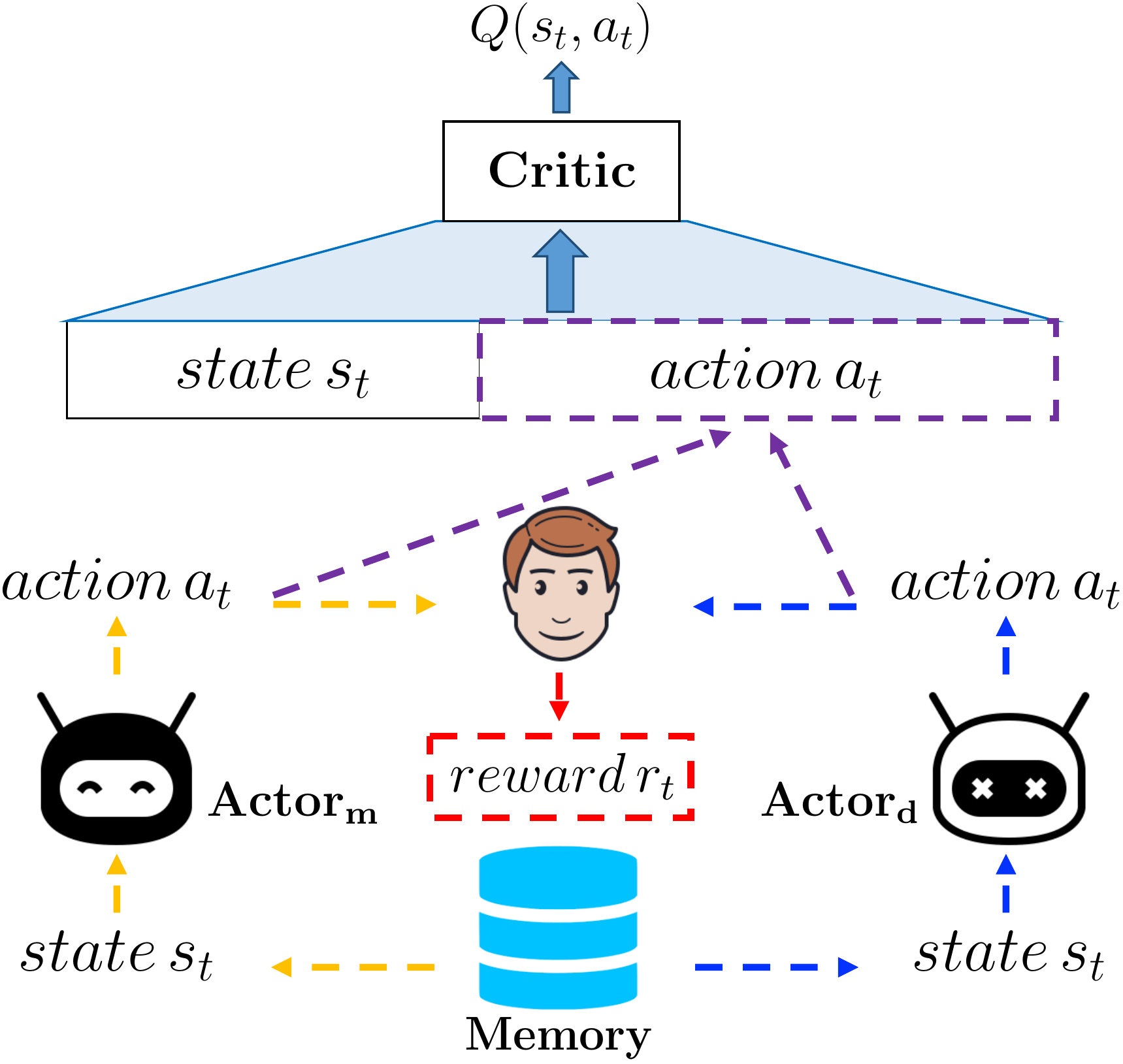}
	\caption{An overview of the proposed framework.\label{fig:AC}}
\end{figure}

\subsection{An Overview of the Proposed Framework}
\label{sec:overview}
The multi-agent RL framework with Actor-Critic architecture is illustrated in Figure~\ref{fig:AC}. In our setting, the proposed framework has two RAs (actors), i.e., $Actor_m$ providing recommendations in the entrance page and $Actor_d$ for the recommendations in the item detail page. Actors aim to generate recommendations according to users' browsing histories (state). As mentioned in Section~\ref{sec:introduction}: (i) the RAs are sequentially activated to interact with users, (ii) the RAs share the same memory of users' historical behavior data (state), and (iii) the RAs will work jointly to maximize the overall performance, which is evaluated by a global action-value function (critic). In other words, a global critic controls all actors to enable them to work collaboratively to optimize the same overall performance. To be specific, the critic takes the current state-action pair as the input, and outputs an action-value to evaluate the long-term future rewards corresponding to the current state and action. Next, we will discuss their architecture.

\subsection{The Actor Architecture}
\label{sec:actor}
The goal of the actors is to suggest recommendations based on users' historical browsing behaviors (state), which should address two challenges: (i) how to capture users' dynamic preference in one recommendation session, and (ii) how to generate recommendations according to the learned users' preference. To tackle these challenges, we develop a two-stage framework, where the first stage (i.e. actor) aims to learn users' dynamic preferences, and the second stage targets to generate recommendations. Note that all actors share the same architecture with different parameters.

\subsubsection{First-stage: Actor to Capture Users' Preference}
\label{sec:encoder}
\begin{figure}[t]
	\centering
	\includegraphics[width=\linewidth]{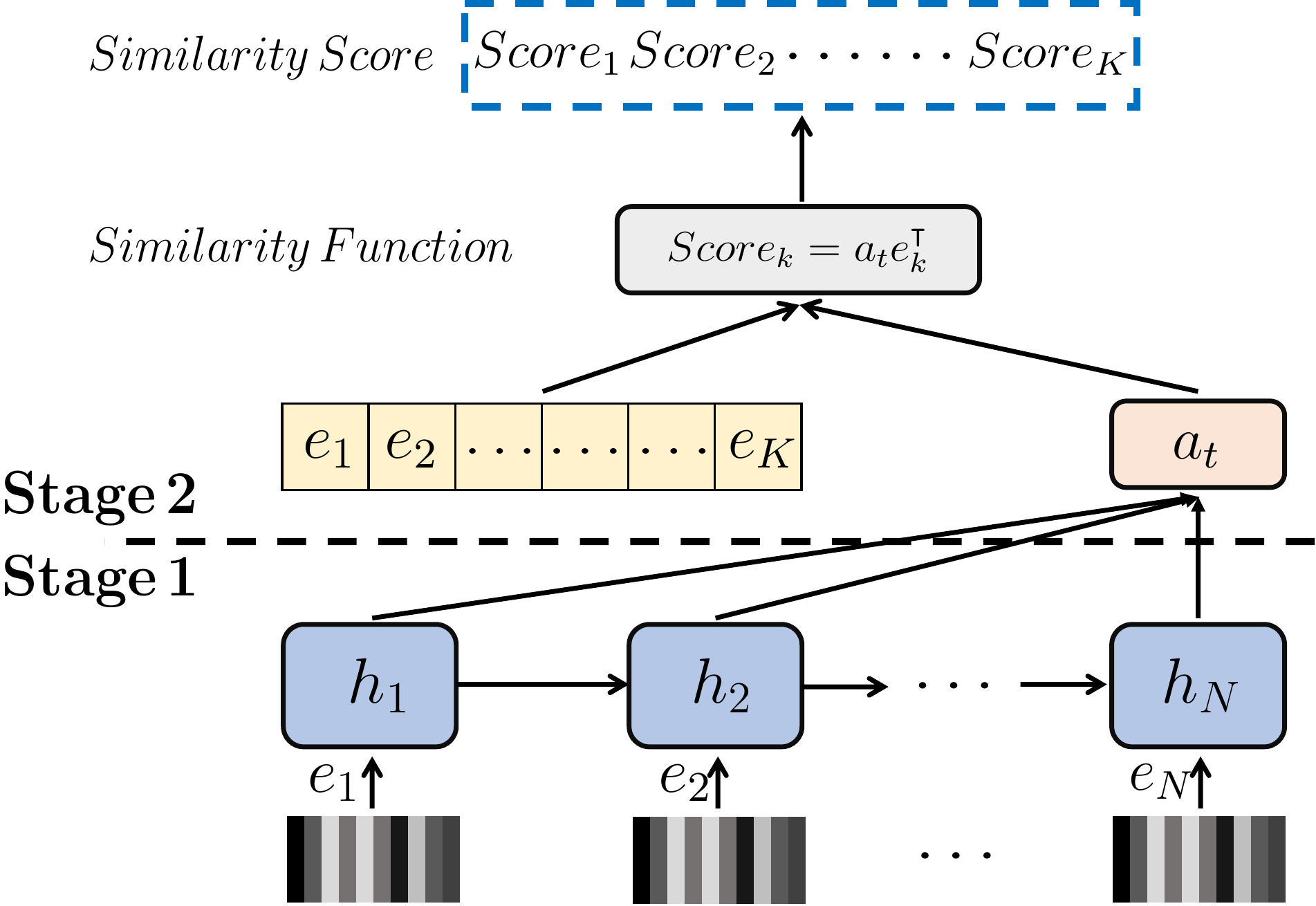}
	\caption{The architecture of the actors.\label{fig:Actor}}
\end{figure}

The sub-figure under the dash line of Figure~\ref{fig:Actor} illustrates the actor architecture that aims to learn users' dynamic preferences during the recommendation session. The actor takes the item representations\footnote{The item representations $\{e_1, \cdots, e_{N}\}$ are dense and low-dimensional vectors, which are pre-trained based on users' browsing history via word embedding~\cite{levy2014neural}, where the clicked items in one recommendation session are treated as a sentence, and each item is treated as a word.} of users' previously clicked or purchased items $\{e_1, \cdots, e_{N}\}$ ($ e_n \in \mathbb{R}^{|E|}$) in sequential order as inputs , and will output the representations of users' dynamic preference in the form of a dense and low-dimensional vector. We introduce a recurrent neural network (RNN) with Gated Recurrent Units (GRU) to capture users' sequential browsing behaviors. To capture user's attention in current session, we introduce an item-level attention mechanism~\cite{li2017neural}, where the actor could dynamically combine the hidden states $\{h_1,\cdots,h_n\}$ of RNN to generate the action $a_t \in \mathbb{R}^{|H|}$.

\subsubsection{Second-stage: Recommendation Generation}
\label{sec:decoder}
The sub-figure above the dash line of Figure~\ref{fig:Actor} illustrates the recommendation generation stage, which targets to generate recommendations according to users' preference learned by the Actor. To be specific, a similarity function between the representations of user's current preference $a_t$ and each candidate item is proposed to calculate a similarity score:
\begin{equation}\label{equ:score_m}
	Score_k =  a_t e_k^\intercal, 
\end{equation}
\noindent Then we select the item with the highest similarity score as the output of the second stage, i.e., the next item to be recommended according to users' current preference. Note that we adopt an item-recall mechanism to reduce the number of relevant items~\footnote{ In general, user's preference in current session should be related to user's previously clicked/purchased items before the current session (say $\mathcal{L}$). Thus for each item in $\mathcal{L}$, we collect a number of most similar items in terms of cosine similarity from the whole item space, and combine all collected items as the initial item space $\mathcal{ I }$ of current recommendation session. During the current session, when a user clicks or purchases an item, we will also add a number of its most similar items into the item space $\mathcal{ I }$.}. Next, we will discuss the architectures of the critic. 

\subsection{The Critic Architecture}
\label{sec:critic}
The RAs introduced in Section~\ref{sec:actor} should work collaboratively to optimize the same overall objective, which is measured by a global critic network (action-value function). In other words, the global critic will control all RAs to work cooperatively to maximize the global performance. Specifically, the input of critic is the current state-action pair $(s_t, a_t)$, and the output is an action-value $Q(s_t, a_t)$ that evaluates the future cumulative rewards starting from the current state and action. According to the $Q(s_t, a_t)$, the actors will update their parameters to generate more accurate recommendations, which will be discussed in the following subsections. The global critic should tackle one challenge, i.e., how to capture user's preferences in different scenarios. In other words, users in different scenarios have different preferences, even though they have the same browsing histories.
\begin{figure}[t]
	\centering
	\includegraphics[width=77.7mm]{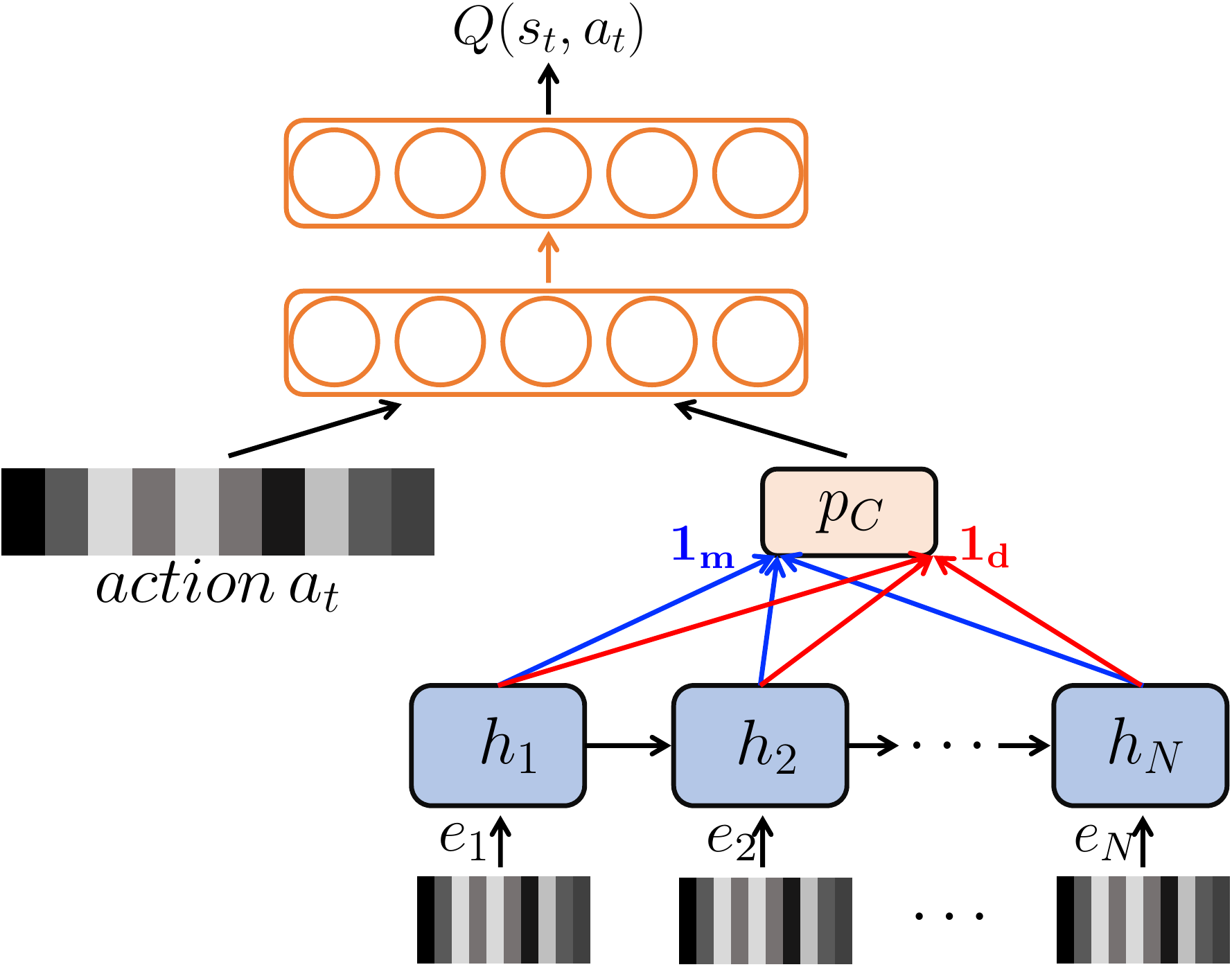}
	\caption{The architecture of the critic.\label{fig:Critic}}
\end{figure}

Figure~\ref{fig:Critic} illustrates the architecture of the global critic, which takes the current state $s_t$ and action $a_t$ as input. We follow the same strategy to feed the item representations of users' previously clicked/purchased items $\{e_1, \cdots, e_{N}\}$ into an RNN with GRU units. Note that the RNNs in actors and global critic share the same architecture with independent parameters. In order to tackle the aforementioned challenge, we introduce two separate item-level attention mechanisms for the entrance page and the item detail page. The intuition is that user's preference in different scenarios is influenced by different parts of her/his browsing histories (i.e. different attention patterns). In other words, we will obtain different representations of user's preferences in different scenarios. The mutually-exclusive indicators $\mathbf{1_m}$ and $\mathbf{1_d}$ will control the activation of two attention mechanisms, i.e., only one attention mechanism is activated each time. Next, we will concatenate the users' preference $p_C$ and current action $a_t$, and feed them into several fully connected layers as a nonlinear approximator to estimate the action-value function $Q(s_t, a_t)$.

\subsection{The Optimization Task}
\label{sec:optimization}
In this subsection, we will discuss the objective functions with the optimization algorithm. As mentioned in Section~\ref{sec:introduction}, the majority of existing model-free RL-based recommender systems need huge amounts of users' behavior data. It is hard to be satisfied in the practical business, because real users will leave the system quickly if they receive almost randomly recommended items, which frequently happens in the initial model training (exploration) stage. Furthermore, since users' skip behaviors (with zero reward) occur much frequently than users' click/purchase behaviors (with positive reward), the distributions of immediate reward function $r_t(s_t,a_t)$ are extremely unbalanced, which can result in an inaccurate update of action-value function. Therefore, we proposed a model-based RL framework for the whole-chain recommendation system, which can approximate the environment (user behaviors) to reduce the desired training data amount and perform more accurate optimization of the action-value function~\cite{brafman2002r,kearns2002near,hu2018reinforcement}. 

Under our setting with two scenarios, i.e., the entrance page and the item detail page, users have three types of behaviors in each scenario. In the entrance page, given a recommended item based on the current state $s_t$, users can: (i) skip the item and continue browsing in the entrance page with a probability $p_m^{s}(s_t,a_t)$, (ii) click the item and go to the item detail page with probability $p_m^{c}(s_t,a_t)$, or (iii) leave the session with probability $p_m^{l}(s_t,a_t)$. Similarly, in the item detail page, given a state-action pair, users can: (i) click the item and go to another item detail page with probability $p_d^{c}(s_t,a_t)$, (ii) skip the item and go back to the entrance page with the probability $p_d^{s}(s_t,a_t)$, or (iii) leave the session with probability $p_d^{l}(s_t,a_t)$. Then the approximation (target) of the action value function, referred as to $y_t$, can be formulated as follows:
\begin{equation}\label{equ:y_t}
	\begin{aligned}
		y_t =&\,\big[p_m^{s}(s_t,a_t) \cdot \gamma Q_{\mu'}(s_{t+1}, \pi'_m(s_{t+1}))\\
		+ &\,\,\,p_m^{c}(s_t,a_t)\cdot \big(r_t+\gamma Q_{\mu'}(s_{t+1}, \pi'_d(s_{t+1}))\big)\\
		+ &\,\,\,p_m^{l}(s_t,a_t)\cdot r_t\big]\mathbf{1_m}\\
		+ &\,\big[p_d^{c}(s_t,a_t) \cdot \big(r_t+\gamma Q_{\mu'}(s_{t+1}, \pi'_d(s_{t+1}))\big)\\
		+ &\,\,\,p_d^{s}(s_t,a_t)\cdot \gamma Q_{\mu'}(s_{t+1}, \pi'_m(s_{t+1}))\\
		+ &\,\,\,p_d^{l}(s_t,a_t)\cdot r_t\big]\mathbf{1_d},
	\end{aligned}
\end{equation}
\noindent where the mutually-exclusive indicators $\mathbf{1_m}$ and $\mathbf{1_d}$ control the activation of two scenarios. Notations $\pi'_m$, $\pi'_d$ and $\mu'$ represent the parameters of the target network of $Actor_m$, $Actor_d$ and Critic respectively of the DDPG framework~\cite{lillicrap2015continuous}. In the Eq (\ref{equ:y_t}), the first row corresponds to the ``skip'' behavior in entrance page that leads to a nonzero Q-value, and the $Actor_m$ will continue recommending next item according to new state $s_{t+1}$; the second row corresponds to the ``click'' behavior in the entrance page that leads to a positive immediate reward and a nonzero Q-value, and $Actor_d$ is activated to recommend next item; the third row corresponds to the ``leave'' behavior in the entrance page that leads to a negative immediate reward, and the session ends; the fourth row corresponds to the ``click'' behavior in the item detail page that leads to a positive immediate reward and a nonzero Q-value, and $Actor_d$ will continue recommending next item; the fifth row corresponds to the ``skip'' behavior in the item detail page that leads to a nonzero Q-value, and $Actor_m$ is re-activated to generate next recommendation; the last row corresponds to the ``leave'' behavior in the item detail page that leads to a negative immediate reward, and the session ends.

We leverage the off-policy DDPG algorithm~\cite{lillicrap2015continuous} to update the parameters of the proposed Actor-Critic framework based on the samples stored in a replay buffer~\cite{mnih2015human}, and we introduce separated evaluation and target networks~\cite{mnih2013playing} to help smooth the learning and avoid the divergence of parameters. Next, we will discuss the optimization of user behavior probabilities, actors and critic.

\subsubsection{Estimating the State Transition Probability}
\label{sec:probabilities}
In fact, user behavior probabilities $[p_m^{c}(s_t,a_t),p_m^{s}(s_t,a_t),p_m^{l}(s_t,a_t)]$ and $[p_d^{c}(s_t,a_t),$  $p_d^{s}(s_t,a_t),p_d^{l}(s_t,a_t)]$ are state transition probabilities introduced in Section~\ref{sec:problem}. In other word, users' different behaviors result in different state transitions. We develop one \textit{probability network}, to estimate the state transition probabilities. The architecture is similar with the critic network shown in Figure~\ref{fig:Critic}, which takes current state-action pair as input, while the only difference is that the output layer has two separate softmax layers that predicts the state transition probabilities of two scenarios. To update the parameters of probability networks, we leverage supervised learning techniques as standard model-based RL, which minimize the \textit{cross entropy} loss between predicted probability vector and ground truth one-hot vector (e.g. $[1,0,0]$ represents ``click'' behavior).

\subsubsection{Optimizating the Critic Parameters}
\label{sec:optimization_critic}
The critic, i.e., the action value function $Q_\mu(s_t, a_t)$, can be optimized by minimizing the loss functions $L(\mu)$ as follows:
\begin{equation}\label{equ:L}
	L(\mu)=\mathbb{E}_{s_t, a_t,r_t,s_{t+1}}\big(y_t-Q_{\mu}(s_t, a_t)\big)^{2},
\end{equation}
\noindent where $\mu$ represents all the parameters of critic (evaluation network), and $y_t$ is defined in Eq (\ref{equ:y_t}). The parameters $\pi'_m$, $\pi'_d$ and $\mu'$ learned from the previous iteration and the state transition probabilities in Eq (\ref{equ:y_t}) are fixed when optimizing the loss function $L(\mu)$. The derivative of the loss function $L(\mu)$ with respective to parameters $\mu$ is presented as follows:
\begin{equation}\label{equ:differentiating}
	\begin{aligned}
		\nabla_{\mu}L(\mu) & =\mathbb{E}_{s_t, a_t,r_t,s_{t+1}}\big[(y_t-Q_{\mu}(s_t, a_t))\nabla_{\mu}Q_{\mu}(s_t, a_t)\big].
	\end{aligned}
\end{equation}
\begin{algorithm}[t]
	\caption{\label{alg:training} Off-policy Training for DeepChain with DDPG.}
	\raggedright
	\begin{algorithmic} [1]
		\STATE Randomly initialize actor and critic networks  $\pi_m$, $\pi_b$, $Q_{{\mu}}$	
		\STATE Initialize target network $\pi'_m\leftarrow\pi_m, \pi'_b\leftarrow\pi_b, Q_{{\mu'}} \leftarrow Q_{{\mu}}$
		\STATE Initialize the capacity of replay buffer $\mathcal{D}$
		\FOR{$session =1:G$}
		\STATE  Receive initial observation state $s_1$
		\FOR{$t=1:T$}
		\STATE  Observe $(s_t, a_t, r_t,s_{t+1})$ following off-policy $b(s_t)$
		\STATE  Store transition $(s_t, a_t,  r_t,s_{t+1})$ in $\mathcal{D}$
		\STATE  Sample minibatch of $\mathcal{N}$ transitions $(s, a, r, s')$ from $\mathcal{D}$
		\STATE Update $[p_m^{c}(s,a),p_m^{s}(s,a),p_m^{l}(s,a)]$ and $[p_d^{c}(s,a),$  $p_d^{s}(s,a),p_d^{l}(s,a)]$ according to Section~\ref{sec:probabilities}
		\STATE  Compute $y$ according to Eq (\ref{equ:y_t})
		\STATE  Update Critic by minimizing $\frac{1}{\mathcal{N}}\sum_{n}\big(y-Q_{\mu}(s, a)\big)^{2}$ according to Eq (\ref{equ:differentiating})
		\STATE  Update Actors $\pi_m$, $\pi_b$ using the sampled policy gradient according to Eq (\ref{equ:policy})
		\STATE Update the target networks: $${\mu'}\leftarrow\tau{\mu}+(1-\tau){\mu'}$$
		$$	{\pi'_m}\leftarrow\tau{\pi_m}+(1-\tau){\pi'_m}	$$ $$	{\pi'_b}\leftarrow\tau{\pi_b}+(1-\tau){\pi'_b}	$$
		\ENDFOR 
		\ENDFOR 
	\end{algorithmic}
\end{algorithm}

\subsubsection{Optimizating the Actor Parameters}
\label{sec:optimization_actor}
The actors can be updated by maximizing $Q_{\mu}(s_t, a_t)$ using the policy gradient:
\begin{equation}\label{equ:policy}
	\nabla_{{\pi}}L(\pi)  \approx  \mathbb{E}_{s_t} \big[\nabla_{a_t}Q_{\mu}(s_t, a_t) \,   \nabla_{{\pi}}{\pi}(s_t)\big],
\end{equation}
\noindent where $\pi$ can represent the parameters of $Actor_m$ or $Actor_d$.

\subsubsection{The Training Algorithm}
\label{sec:algorithm}
The off-policy training algorithm for DeepChain is presented in Algorithm~\ref{alg:training}. In each iteration, there are two stages, i.e., 1) transition generating stage (lines 7-8), and 2) parameter updating stage (lines 9-15). For transition generating stage: we first observe the transition $(s_t, a_t,  r_t,s_{t+1})$ following offline behavior policy $b(s_t)$ that generates the historical behavior data (line 7), then we store the transition $(s_t,a_t,r_t,s_{t+1})$ into the replay buffer $\mathcal{D}$ (line 8). For parameter updating stage: we first sample mini-batch of transitions $(s, a, r, s')$ from $\mathcal{D}$ (line 9), then we update the state transition probabilities by supervised learning techniques as mentioned in Section~\ref{sec:probabilities} (line 10), and finally we update critic and actors (lines 11-14) following a standard DDPG procedure~\cite{lillicrap2015continuous}. Note that it is straightforward to extend the off-policy training to on-policy training: in transition generating stage, we can collect transitions $(s_t, a_t,  r_t,s_{t+1})$ with exploration during the interactions between RAs and real users.

\subsection{The Test Tasks}
\label{sec:test}
In this subsection, we will present the test tasks of the DeepChain framework. We propose two test tasks, i.e., (i) Offline test: testing the proposed framework based on user's historical behavior data; and (ii) Online test: testing the proposed framework in a real online environment where the RAs interact with real-world users and receive an immediate reward (real-time feedback) of the recommended items from users. Note that offline test is necessary because recommendation algorithms should be pre-trained (by the off-policy algorithm in Section~\ref{sec:optimization}) and evaluated offline before launching in the real online system, which ensures the recommendation quality and mitigates the negative influence on user experience.  

\begin{algorithm}[t]
	\caption{\label{alg:test_on} Online Test of DeepChain.}
	\raggedright
	\begin{algorithmic} [1]
		\STATE Initialize actors with well trained parameters ${\pi_m}$ and ${\pi_b}$
		\STATE  Observe initial the state $s_1$
		\FOR{$t=1:T$}
		\IF{the user in main page}
		\STATE  Execute an action $a_t$ following policy $\pi_m(s_t)$
		\ELSE 
		\STATE Execute an action $a_t$ following policy $\pi_b(s_t)$
		\ENDIF
		\STATE  Observe the reward $r_t$and transition to new state $s_{t+1}$
		\ENDFOR 
	\end{algorithmic}
\end{algorithm}

\subsubsection{Online Test}
\label{sec:on_test}
The online test algorithm is presented in Algorithm~\ref{alg:test_on}. In each iteration of a recommendation session, given the current state $s_t$ and scenario, one actor is activated to recommend an item to user following policy ${\pi_m}$ or ${\pi_b}$ (line 5 or 7). Then the system observes the reward $r_t$ from user and updates the state to $s_{t+1}$ (line 9). 

\begin{algorithm}
	\caption{\label{alg:test_off}Offline Test of DeepChain.}
	\raggedright
	{\bf Input}: Item list $\mathcal{I} = \{i_1, \cdots, i_N\}$ and related reward list $\mathcal{R} = \{r_1, \cdots, r_N\}$  of a session.\\
	{\bf Output}:Re-ranked recommendation list $\mathcal{L}$\\
	\begin{algorithmic} [1]
		\STATE Initialize actor with well trained parameters ${\pi}$
		\STATE  Receive initial observation state $s_1$
		\WHILE{$|\mathcal{I}| > 0$}
		\STATE  Execute an action $a_t$ following policy $\pi(s_t)$
		\STATE  Add $a_t$ into the end of $\mathcal{L}$
		\STATE  Observe reward $r_t$ from users (historical data)
		\STATE  Observe new state $s_{t+1}$
		\STATE  Remove $a_t$ from $\mathcal{I}$ 
		\ENDWHILE
	\end{algorithmic}
\end{algorithm}

\begin{table*}[]
	\centering
	\caption{Performance comparison of different recommender systems}
	\label{table:result1}
	\begin{tabular}{|c|c|c|c|c|c|c|c|}
		\hline
		\multirow{2}{*}{Scenarios} & \multirow{2}{*}{Metrics} & \multicolumn{6}{c|}{Algorithms} \\ \cline{3-8} 
		&                   & W\&D  & DFM  & GRU  & DDPG  &  MA &  DeepChain \\ \hline\hline
		\multirow{3}{*}{\begin{tabular}[c]{@{}c@{}}Entrance\\ Page \end{tabular}} & MAP & 0.106 & 0.108 & 0.113 & 0.117 &  0.121 &  \textbf{0.126} \\ \cline{2-8} 
		& improv.(\%) & 18.87  & 16.67  & 11.50  & 7.693  & 4.132  & -  \\ \cline{2-8} 
		& p-value & 0.000  & 0.000  & 0.000  & 0.000  & 0.003  & -  \\ \hline\hline
		\multirow{3}{*}{\begin{tabular}[c]{@{}c@{}}Entrance\\ Page \end{tabular}} & NDCG@40 & 0.189 & 0.193 & 0.201 & 0.209 & 0.215 & \textbf{0.225} \\ \cline{2-8} 
		& improv.(\%) & 19.05  & 16.58  & 11.95  & 7.656  & 4.651  & -  \\ \cline{2-8} 
		& p-value & 0.000  & 0.000  & 0.000  & 0.000  & 0.003  & -  \\ \hline\hline
		\multirow{3}{*}{\begin{tabular}[c]{@{}c@{}c@{}}Item\\ Detail\\ Page \end{tabular}} & MAP & 0.081 & 0.083 & 0.086 & 0.090 & 0.093 & \textbf{0.096}  \\ \cline{2-8} 
		& improv.(\%) & 18.52  & 15.66  & 11.63  & 6.667  & 3.226  &  - \\ \cline{2-8} 
		& p-value & 0.000  & 0.000  & 0.000  & 0.000  & 0.006  & -  \\ \hline\hline
		\multirow{3}{*}{\begin{tabular}[c]{@{}c@{}c@{}}Item\\ Detail\\ Page \end{tabular}} & NDCG@40 & 0.166 & 0.169 & 0.176 & 0.183 & 0.190 & \textbf{0.197}  \\ \cline{2-8} 
		& improv.(\%) & 18.67  & 16.57  & 11.93  & 7.650  & 3.684  &  - \\ \cline{2-8} 
		& p-value & 0.000  & 0.000  & 0.000  & 0.000  & 0.005  & -  \\ \hline
	\end{tabular}
\end{table*}

\subsubsection{Offline Test}
\label{sec:off_test}
The intuition of offline test is that, given historical offline recommendation session data, if DeepChain works well, it can re-rank the items in this session and the ground truth clicked items can be sorted at the top of the new list. The DeepChain only re-ranks items in this session rather than all items from item space, because we only know the ground truth rewards corresponding to the existing items of this session in the offline data. The offline test algorithm is presented in Algorithm~\ref{alg:test_off}. In each iteration, given $s_t$, the actor recommends an item to user following policy ${\pi}$ (line 4), where we calculate the recommending score of all items in the item list $\mathcal{I}$, and select the one with the highest score. Then we add the selected item into a new recommendation list $\mathcal{L}$ (line 5), and record reward $r_t$ from historical data (line 6). Next we update the state to $s_{t+1}$ (line 7). Finally, we remove the selected item from $\mathcal{I}$ (line 8), which avoids repeatedly recommending the same items. Note that in an offline test setting, we collect user behavior data in two scenarios and re-rank the items in each scenario.
\section{Experiment}
\label{sec:experiments}
In this section, we conduct extensive experiments to evaluate the effectiveness of the proposed framework based on a real e-commerce platform. We mainly focus on two questions: (1) how the proposed framework performs compared to the state-of-the-art baselines; and (2) how the components in the framework contribute to the performance. We first introduce experimental settings. Then we seek answers to the above two questions. Finally, we study the impact of key parameters on the performance of the proposed framework.

\subsection{Experimental Settings}
\label{sec:experimental_settings}
We collected a real-world dataset of December 2018 from an e-commerce platform, JD.com. We randomly collect 500,000 recommendation sessions (with 19,667,665 items) in temporal order, and leverage the first 80\% sessions as the training/validation datasets and the later 20\% sessions as the (offline) test dataset. For a new session, the initial state $s_1$ is $N = 50$ previously clicked or purchased items obtained from users' previous sessions~\cite{guo2020joint}. The immediate reward $r_t$ of click/skip/leave behavior is empirically set as 1, 0, and -2, respectively. The dimensions of item representation vector and hidden state of RNN are $|E| = 20$ and $|H| = 64$. The discounted factor $\gamma = 0.95$, and the soft update rate for target networks is $\tau = 0.01$. We select the parameters of the DeepChain framework via cross-validation, and do parameter-tuning for baselines for a fair comparison. More details about parameter analysis will be discussed in the following subsections. For offline test, we select \textbf{NDCG}~\cite{jarvelin2002cumulated} and \textbf{MAP}~\cite{turpin2006user} as metrics. For online test, we use the overall reward in one recommendation session as the metric. 

\subsection{Performance Comparison for Offline Test}
\label{sec:ev_overall}
We compare the proposed framework with the following representative baseline methods:

\begin{itemize}[leftmargin=*]
	\item \textbf{W\&D} \cite{cheng2016wide}: This baseline is a wide \& deep model for jointly training feed-forward neural networks with embeddings and linear model with feature transformations for generic recommender systems with sparse inputs. 
	\item \textbf{DFM} \cite{guo2017deepfm}: DeepFM is a deep neural network model that integrates the architectures of factorization-machine (FM) and wide \& deep model. It models low-order feature interactions like FM and models high-order feature interactions like W\&D. 
	\item \textbf{GRU}~\cite{hidasi2015session}: GRU4Rec leverages the RNN with GRU to predict what a user will click next based on the clicking history. We also keep $N = 50$ clicked items as the state for a fair comparison.
	\item \textbf{DDPG}~\cite{dulac2015deep}: This baseline is the conventional Deep Deterministic Policy Gradient. The input for Actor is the concatenation of embeddings of users' historical clicked items (state). The input for Critic is the concatenation of the state and a recommended item (action).
	\item \textbf{MA}~\cite{lowe2017multi}: This is a multi-agent model-free RL model, where agents learn a centralized critic based on the observations and actions of all agents. 
\end{itemize}

The results are shown in Table~\ref{table:result1}. Note that in the offline test, we separately collect user behavior data from two scenarios and re-rank the items in each scenario by the corresponding RA. We make the following observations:
(i) The DFM achieves better performance than W\&D, where DeepFM can be trained end-to-end without any feature engineering, and its wide part and deep part share the same input and also the embedding vector.
(ii) GRU outperforms W\&D and DFM, since GRU can capture the temporal sequence within one recommendation session, while W\&D and DFM neglect it. 
(iii) DDPG achieves better performance than GRU, since DDPG can optimize the overall performance of one recommendation session, but GRU only maximizes the immediate reward for next recommended item. This result validates the advantage of RL techniques in recommendations.
(iv) DDPG performs worse than MA and DeepChain, because DDPG trains each RA in each scenario separately, while MA and DeepChain are multi-agent models where RAs are jointly trained on two scenarios (the whole dataset) to optimize the global performance.
(v) DeepChain outperforms MA, since model-based RL model like DeepChain can perform more accurate optimization of the action-value function based on less training data.
To sum up, DeepChain outperforms representative baselines, which demonstrates its effectiveness in recommendations.

\begin{figure}[H]
	\centering
	\includegraphics[width=\linewidth]{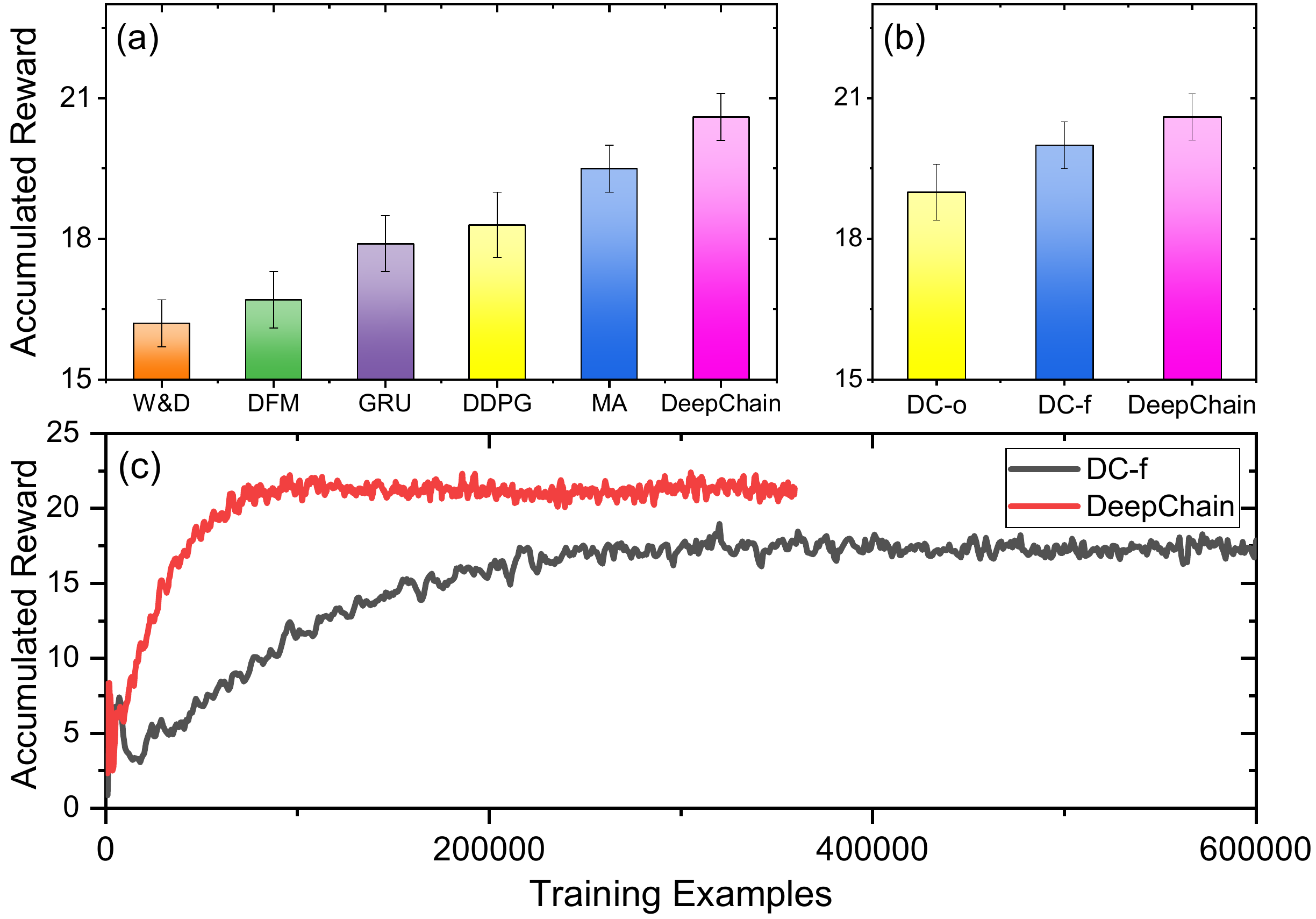}
	\caption{Overall performance comparison in online test.}
	\label{fig:architecture}
\end{figure}

\subsection{Performance Comparison for Online Test}
Following~\cite{zhao2018recommendations,zhao2018deep}, we evaluate the proposed framework on a simulated online environment, where a deep neural network is pre-trained to estimate users' behaviors based on historical state-action pairs. Furthermore, to answer the second question, we systematically eliminate the corresponding components of DeepChain by defining the following two variants:

\begin{itemize}[leftmargin=*]
	\item \textbf{DC-o}: This variant is a one-agent version of DeepChain. In other words, only one RA is trained to generate recommendations on both the entrance page and item detail page.
	\item \textbf{DC-f}: This variant is a model-free version of DeepChain, which does not estimate the user behavior probabilities as mentioned in Section~\ref{sec:probabilities}.
\end{itemize}

The results are shown in Figure~\ref{fig:architecture}. It can be observed: 
(i) We observe similar online test comparison results between DeepChain and the state-of-the-art baselines as these in the offline test in Figure~\ref{fig:architecture} (a).
(ii) DC-o performs worse than DeepChain, since DC-o only trains one RA for both two scenarios. This result indicates that users' interests in different scenarios are different. Thus developing separate RAs for different scenarios is necessary.
(iii) DC-f achieves worse performance than DeepChain. The key reasons include: (iii-a) model-free version DC-f requires more training data, as illustrated in Figure~\ref{fig:architecture}(c), DC-f converges much slower than DeepChain; and (iii-b) DC-f performs less accurate optimization of $Q(s_t,a_t)$ than model-based model DeepChain. This result validates the effectiveness of model-based RL in recommendations.
In summary, appropriately developing separate RAs and introducing model-based techniques to update action-value function can boost the recommendation performance. 

\begin{figure}[H]
	\centering
	\includegraphics[width=\linewidth]{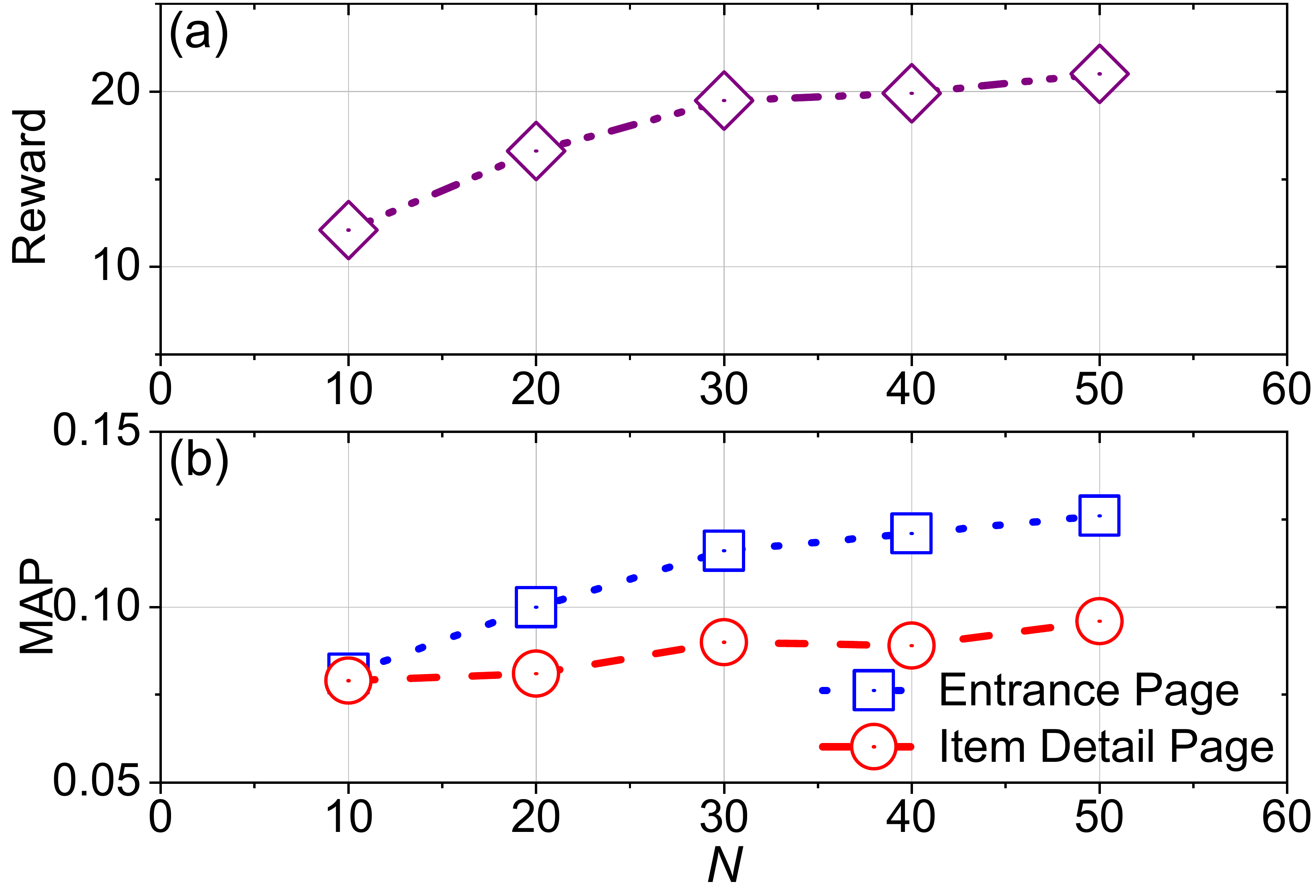}
	\caption{Parameter sensitiveness.}
	\label{fig:parameters}
\end{figure}

\subsection{Parameter Sensitivity Analysis}
\label{sec:parametric}
In this subsection, we investigate how the proposed framework DeepChain performs with the changes of $N$, i.e., the length of users' browsing history (state), while fixing other parameters.

We can observe: (i) Figure \ref{fig:parameters} (a) demonstrates the parameter sensitivity of $N$ in online test. We can find that with the increase of $N$, the overall performance improves. 
(ii) Figure \ref{fig:parameters} (b) shows the parameter sensitivity of $N$ in the offline test task. We can observe that the recommendation performance of the entrance page is more sensitive with the increase of $N$ than that of the item detail page. The reason is that users' interests are different in two scenarios: in the entrance page, users' preferences are diverse, thus including longer browsing history can better discover users' various interests; while in one specific item's detail page, users' attention mainly focuses on the similar items to this specific item, in other words, users would like to compare this item with similar ones, thus involving longer browsing history cannot significantly improve the performance.
\section{Related Work}
\label{sec:related_work}
In this section, we briefly review works related to our study, i.e., RL-based recommender systems~\cite{zhao2019deep,zhang2020deep}. Practical recommender systems are always with millions of items (discrete actions) to recommend. Thus, most RL-based models will become inefficient since they are not able to handle such a large discrete action space. A Deep Deterministic Policy Gradient (DDPG) algorithm is introduced to mitigate the large action space issue in practical RL-based recommender systems~\cite{dulac2015deep}. To avoid the inconsistency of DDPG and improve recommendation performance, a tree-structured policy gradient is proposed in~\cite{chen2018large}. Biclustering technique is also introduced to model recommender systems as grid-world games so as to reduce the state/action space~\cite{choi2018reinforcement}. To solve the unstable reward distribution problem in dynamic recommendation environments, approximate regretted reward technique is proposed with Double DQN to obtain a reference baseline from individual customer sample~\cite{chen2018stabilizing}. Users' positive and negative feedback, i.e., purchase/click and skip behaviors, are jointly considered in one framework to boost recommendations, since both types of feedback can represent part of users' preference~\cite{zhao2018recommendations}. 
A page-wise recommendation framework is proposed to jointly recommend a page of items and display them within a 2-D page~\cite{zhao2017deep,zhao2018deep}. CNN technique is introduced to capture the item display patterns and users' feedback of each item in the page. In the news feed scenario, a DQN based framework is proposed to handle the challenges of conventional models, i.e., (1) only modeling current reward like CTR, (2) not considering click/skip labels, and (3) feeding similar news to users~\cite{zheng2018drn}. An RL framework for explainable recommendation is proposed in~\cite{wang2018reinforcement}, which can explain any recommendation model and can flexibly control the explanation quality based on the application scenario. A policy gradient-based top-K recommender system for YouTube is developed in~\cite{chen2018top}, which addresses biases in logged data through incorporating a learned logging policy and a novel top-K off-policy correction. In~\cite{feng2018learning}, multi-scenario ranking is formulated as a fully cooperative, partially observable, multi-agent sequential decision problem, denoted as MA-RDPG. While similar in motivation to our approach, it is on the top of model-free approach, which requires a larger amount of user-agent interaction data and suffers from inaccurate update of the action-value function. Other RL-based recommendation applications include sellers' impression allocation~\cite{cai2018reinforcement}, and fraudulent behavior detection~\cite{cai2018reinforcement1}.
\section{Conclusion}
\label{sec:conclusion}
In this paper, we propose a novel multi-agent model-based reinforcement learning framework (DeepChain) for the whole-chain recommendation problem. It is able to collaboratively train multiple RAs for different scenarios by a model-based optimization algorithm. Multi-agent RL-based recommender systems have three advantages: (i) the RAs are sequentially activated to capture the sequential dependency of users' behaviors among different scenarios; (ii) the RAs share the same memory of users' historical behavior information to make more accurate decisions, and (iii) the RAs will work collaboratively to maximize the global performance of one recommendation session. Note that we design a model-based RL optimization algorithm that can reduce the requirement of training data and perform more accurate optimization of the action-value function than model-free algorithms. We conduct extensive experiments based on a real-world dataset from an e-commerce platform. The results show that (i) DeepChain can significantly enhance the recommendation performance; and (ii) multi-agent techniques and model-based RL can enhance the recommendation task.

\section*{ACKNOWLEDGEMENTS}
This work is supported by National Science Foundation (NSF) under grant numbers IIS1907704, IIS1928278, IIS1714741, IIS1715940, IIS1845081 and CNS1815636.

\bibliographystyle{ACM-Reference-Format}
\bibliography{9Reference} 
\end{document}